\documentclass[10pt,a4paper]{article}
\usepackage{amsmath,amssymb,amsfonts}
\usepackage{graphicx}
\usepackage{hyperref}
\usepackage[margin=0.75in]{geometry}
\usepackage{float}
\usepackage{bm}
\usepackage{booktabs}
\usepackage{tabularx}
\usepackage{subcaption}

\usepackage{setspace}
\usepackage{titlesec}
\titlespacing*{\section}{0pt}{8pt}{4pt}
\titlespacing*{\subsection}{0pt}{6pt}{3pt}

\AtBeginDocument{
  \setlength{\abovedisplayskip}{4pt}
  \setlength{\belowdisplayskip}{4pt}
  \setlength{\abovedisplayshortskip}{2pt}
  \setlength{\belowdisplayshortskip}{2pt}
}

\let\oldbibliography\thebibliography
\renewcommand{\thebibliography}[1]{%
  \oldbibliography{#1}%
  \setlength{\itemsep}{0pt}%
  \setlength{\parskip}{0pt}%
}

\title{Causal Attribution of Coastal Water Clarity Degradation to Nickel Processing Expansion at the Indonesia Morowali Industrial Park, Sulawesi}

\author{
Sandy H. S. Herho$^{1,2,3,*}$,
Alfita P. Handayani$^{3,4}$,
Iwan P. Anwar$^{5}$,\\
Faruq Khadami$^{5}$,
Karina A. Sujatmiko$^{5}$,
Doandy Y. Wibisono$^{6,7}$,\\
Rusmawan Suwarman$^{8}$,
and Dasapta E. Irawan$^{9}$
}

\date{}

\begin{document}
\maketitle

\begin{center}
\small
$^{1}$Department of Earth and Planetary Sciences, University of California,
Riverside, CA 92521, USA\\
$^{2}$School of Systems Science and Industrial Engineering,
State University of New York, Binghamton, NY 13902, USA\\
$^{3}$Center for Agrarian Studies, Bandung Institute of Technology,
Bandung, West Java 40132, Indonesia\\
$^{4}$Spatial Systems and Cadaster Research Group,
Bandung Institute of Technology, Bandung, West Java 40132, Indonesia\\
$^{5}$Applied and Environmental Oceanography Research Group,
Bandung Institute of Technology, Bandung, West Java 40132, Indonesia\\
$^{6}$Department of Civil and Environmental Engineering,
Colorado School of Mines, Golden, CO 80401, USA\\
$^{7}$Brierley Associates, Englewood, CO 80110, USA\\
$^{8}$Atmospheric Science Research Group, Bandung Institute of Technology,
Bandung, West Java 40132, Indonesia\\
$^{9}$Applied Geology Research Group, Bandung Institute of Technology,
Bandung, West Java 40132, Indonesia\\
$^{*}$Correspondence: sh001@ucr.edu
\end{center}

\begin{abstract}
\noindent
Indonesia's nickel ore export ban has driven rapid expansion of smelting and hydrometallurgical processing capacity at the Indonesia Morowali Industrial Park (IMIP), now the world's largest integrated nickel processing complex, on the coast of Central Sulawesi. Whether this industrialization has degraded the adjacent marine environment remains unquantified. We apply Bayesian structural time-series (BSTS) causal inference to a multi-decadal, multi-sensor satellite ocean color record of the diffuse attenuation coefficient at 490~nm, $K_d(490)$, to test for a causal link between IMIP expansion and nearshore turbidity change. A consensus structural breakpoint, a significant posterior causal effect estimated against a Banda Sea counterfactual, and a distribution-free placebo rank test collectively establish that coastal water clarity deteriorated after the transition from initial nickel pig iron production to hyper-expansion of high-pressure acid leaching facilities for battery-grade nickel. Satellite-derived land cover analysis independently corroborates this timing, showing substantial built-area growth and concurrent tree cover loss within the IMIP footprint. The resulting euphotic zone shoaling occurs in oligotrophic waters supporting high marine biodiversity, where even moderate optical degradation may impair coral photosynthesis and compress depth-dependent reef habitat. These findings quantify a marine environmental cost absent from Indonesia's mineral downstreaming policy discourse and demonstrate a transferable, satellite-based quasi-experimental framework for causal impact assessment at coastal industrial sites in data-limited tropical settings.
\end{abstract}


\section{Introduction}
\label{sec:introduction}

The global transition to low-carbon energy systems has produced a paradox at the mineral frontier: the technologies intended to mitigate climate change (electric-vehicle batteries, grid-scale storage, wind turbines) demand massive quantities of metals, yet extraction of those metals inflicts severe, often poorly quantified, environmental harm. Nickel is a case in point. Global demand is projected to rise substantially by mid-century under decarbonization scenarios~\cite{watari2018}, and battery-grade nickel sulfate has emerged as a critical bottleneck for cathode manufacturing. Indonesia now supplies roughly half of the world's mined nickel from laterite deposits concentrated on Sulawesi and Halmahera~\cite{vanderent2013,mudd2010} and has responded with an aggressive mineral downstreaming policy built on progressive ore export bans that compel domestic processing~\cite{lahadalia2024nickel,Astuti2025}. The resulting industrial expansion has been extraordinary: since 2015, the Indonesia Morowali Industrial Park (IMIP) on the southeastern coast of Sulawesi has grown from a greenfield site to the world's largest integrated nickel processing complex. The marine environmental consequences of this transformation, however, remain virtually undocumented.

Laterite nickel mining is inherently land-intensive. Ore bodies form as shallow, tabular weathering horizons over ultramafic bedrock and require open-pit extraction that strips vegetation and topsoil across wide tropical forest tracts~\cite{vanderent2013}. Empirical measurements show that Indonesia's nickel land footprint has grown dramatically, with disturbance intensities far exceeding those assumed in life-cycle inventories~\cite{heijlen2024}. A counterfactual impact evaluation across Sulawesi villages found that deforestation rose markedly in nickel-mining communities while environmental well-being declined relative to non-mining controls~\cite{lo2024}. These terrestrial impacts (deforestation, soil exposure, generation of fine-grained, metal-enriched processing waste) carry direct implications for coastal water quality through enhanced sediment runoff, tailings discharge, and destruction of riparian buffering. In New Caledonia, prolonged laterite mining has substantially increased terrigenous sediment delivery to the coastal lagoon~\cite{dumas2010} and raised dissolved trace-metal concentrations to ecologically significant levels~\cite{hedouin2007}. No comparable assessment exists for Indonesia despite the far greater pace and scale of its current expansion.

The waters adjacent to IMIP lie within the Coral Triangle, the global epicenter of marine biodiversity. This region harbors the majority of the world's known coral species~\cite{veron2009} and supports fisheries on which millions of coastal livelihoods depend~\cite{hicks2019}. Turbidity ranks among the most ecologically consequential stressors in this setting: elevated suspended sediment attenuates downwelling irradiance, compresses the euphotic zone, and impairs coral photosynthesis, calcification, and recruitment at chronic concentrations well below acute mortality thresholds~\cite{fabricius2005,jones2015}. The satellite-derived diffuse attenuation coefficient at 490~nm, $K_d(490)$, provides a synoptic, temporally continuous proxy for water clarity retrievable from merged multi-sensor ocean color records spanning more than two decades~\cite{doron2007,lee2007}. Detecting an anthropogenic signal against the substantial natural variability driven by monsoons, the El Ni\~{n}o--Southern Oscillation (ENSO), and the Indian Ocean Dipole (IOD) requires methods that go beyond trend analysis to establish causal attribution. Bayesian structural time-series (BSTS) modeling, originally developed for market intervention analysis~\cite{brodersen2015}, provides a rigorous quasi-experimental framework for this purpose: by constructing a synthetic counterfactual from control-zone covariates unaffected by the intervention, BSTS estimates the posterior probability that an observed change reflects local forcing rather than regional climatic variability. Combined with consensus changepoint detection and satellite-based land use and land cover (LULC) intensity analysis~\cite{aldwaik2012}, this approach yields a multi-line-of-evidence attribution methodology that is transparent and scalable across the Indonesian archipelago.

Here we test the hypothesis that rapid industrialization of the Morowali coastline has produced a statistically detectable and causally attributable degradation of nearshore water clarity. Specifically, we (i)~characterize the 1998--2024 climatology of $K_d(490)$ in both the IMIP impact zone and a Banda Sea control zone; (ii)~apply a consensus of multiple changepoint detection algorithms to identify structural breaks in the impact-zone time series; (iii)~use BSTS modeling with control-zone covariates to estimate the causal effect of IMIP operations on $K_d(490)$ and evaluate its significance via a distribution-free placebo rank test; and (iv)~quantify concurrent LULC changes within the IMIP footprint at 10~m resolution. To our knowledge, the results constitute the first satellite-derived causal evidence linking a specific industrial installation in Indonesia to measurable coastal optical degradation.

\section{Data}
\label{sec:data}

\begin{figure}[H]
\centering
\includegraphics[width=\textwidth]{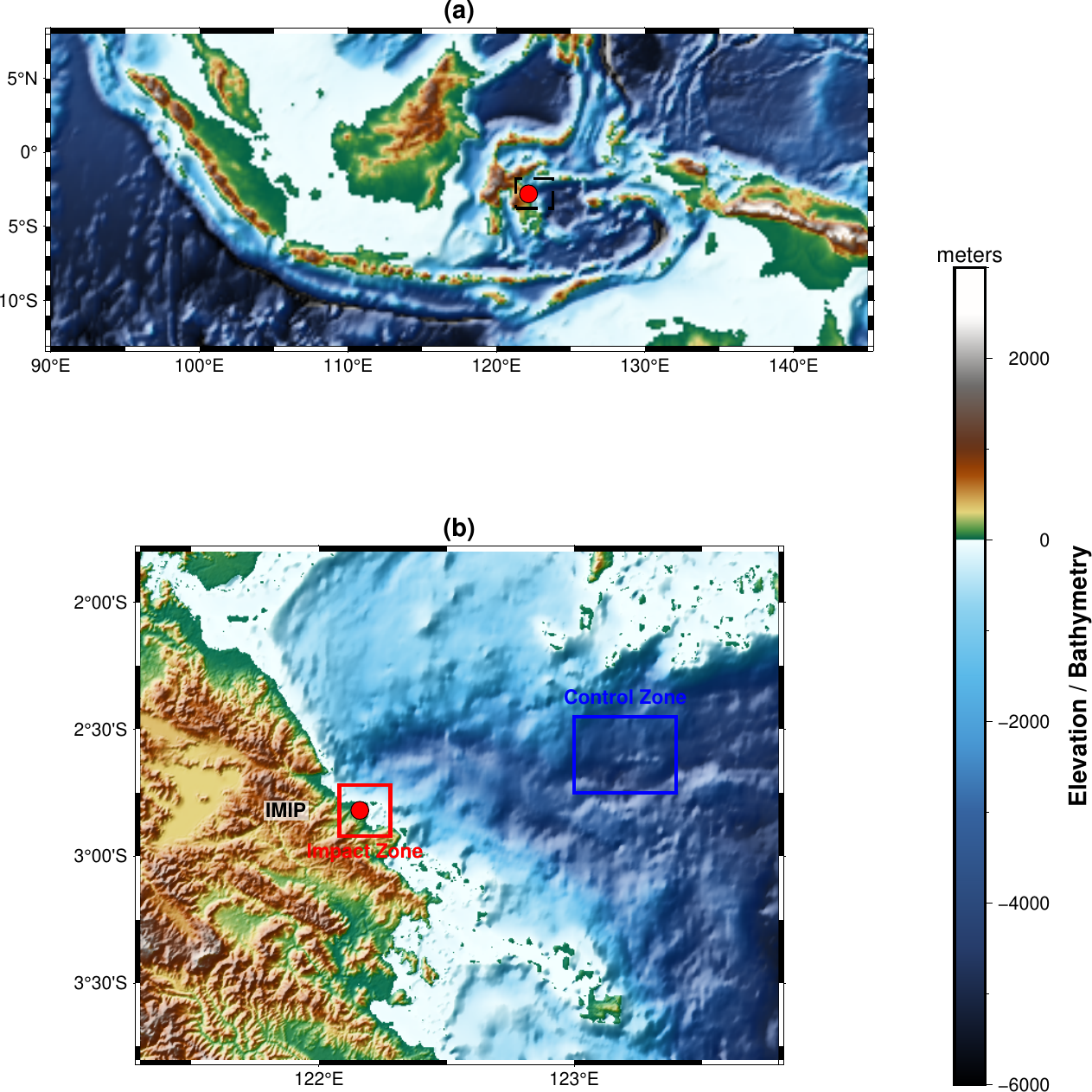}
\caption{Geospatial context of the study domain. (\textbf{a})~Location of IMIP (red circle) within the Indonesian archipelago; the dashed rectangle delineates the study area. (\textbf{b})~Tolo Bay and the western Banda Sea at 15 arc-second bathymetric resolution (SRTM15+V2), showing the impact zone (red box) directly offshore of IMIP and the control zone (blue box) in the open Banda Sea. The color scale denotes elevation and bathymetry in meters.}
\label{fig:studyarea}
\end{figure}

The study domain covers the coastal waters of Morowali Regency, Central Sulawesi, Indonesia, spanning the inner Tolo Bay and western Banda Sea margin within $121.30$--$123.80^{\circ}$E, $1.80$--$3.80^{\circ}$S (Figure~\ref{fig:studyarea}). IMIP, a vertically integrated nickel ore processing complex located at approximately $122.16^{\circ}$E, $2.82^{\circ}$S, sits at the center of the domain. Smelting operations began in April 2015 and underwent rapid capacity expansion after the January 2020 nickel ore export ban~\cite{lahadalia2024nickel,Astuti2025}. Two non-overlapping spatial domains are defined for paired comparison. The \emph{impact zone} ($122.08$--$122.28^{\circ}$E, $2.72$--$2.92^{\circ}$S; ${\sim}256$~km$^{2}$) covers the nearshore waters immediately offshore of the IMIP coastline, where steep lateritic terrain transitions abruptly to a deep continental shelf in a configuration that maximizes land-to-sea sediment delivery. The \emph{control zone} ($123.00$--$123.40^{\circ}$E, $2.45$--$2.75^{\circ}$S; ${\sim}784$~km$^{2}$) occupies open Banda Sea waters 100--150~km from any industrial coastline. This zone experiences the same basin-scale oceanographic and atmospheric forcing but is free of local anthropogenic perturbation. Bathymetric and topographic characterization (Figure~\ref{fig:studyarea}) was derived from the SRTM15+V2 global relief model at 15 arc-second resolution~\cite{tozer2019}, rendered through PyGMT~\cite{uieda2024,wessel2019}.

Terrestrial LULC dynamics were quantified from the Esri 10~m Annual Land Use Land Cover product~\cite{karra2021}, a globally consistent classification derived from Sentinel-2 imagery at 10~m resolution~\cite{drusch2012}. The product applies a deep learning model to assign each cloud-free pixel to one of ten LULC classes. Annual maps were extracted for eight consecutive years (2017--2024) over the impact zone, yielding $2226 \times 2226$ pixel rasters (${\sim}478$~km$^{2}$ valid coverage per year). Water and cloud-flagged pixels were masked, retaining $J = 6$ terrestrial classes (trees, flooded vegetation, crops, built area, bare ground, rangeland) across seven consecutive annual transition intervals for subsequent intensity analysis~\cite{aldwaik2012}.

Ocean color observations of $K_d(490)$ were obtained from the Copernicus Marine Service (CMEMS) GlobColour multi-sensor merged product~\cite{garver1997}. This product merges inter-calibrated water-leaving radiance retrievals from six missions (SeaWiFS, MODIS-Aqua, MERIS, VIIRS-SNPP, VIIRS-JPSS1, and OLCI aboard Sentinel-3A/3B) into a single temporally continuous, gap-filled monthly composite at 4~km resolution from September 1997 onward~\cite{gohin2002,hu2012,xi2021}. The retrieval follows the semi-analytical formulation of Doron et al.~\cite{doron2007}, relating $K_d(490)$ to inherent optical properties estimated through radiative transfer inversion of remote-sensing reflectance. Physically, $K_d(490)$ quantifies the exponential attenuation rate of downwelling irradiance at 490~nm with depth and receives additive contributions from pure seawater absorption, non-algal particles, phytoplankton pigments, and colored dissolved organic matter~\cite{kirk2011,mobley1994}. A total of $n = 324$ monthly fields (January 1998 to December 2024) were extracted over each domain. Area-weighted zonal means were computed using cosine-of-latitude weighting, yielding 16 valid ocean pixels for the impact zone, 49 for the control zone, and 1467 of 2304 for the full domain.

Sea surface temperature (SST) and sea surface salinity (SSS) were sourced from the GLORYS12V1 global ocean eddy-resolving reanalysis at $1/12^{\circ}$ horizontal resolution~\cite{lellouche2021}. The dynamical core is the NEMO ocean general circulation model~\cite{madec2017}, forced by ERA-Interim (prior to 2019) and ERA5 atmospheric boundary conditions, with ocean observations assimilated through a reduced-order Kalman filter~\cite{lellouche2021}. Monthly mean surface potential temperature and practical salinity were extracted over the same domains and temporal span as $K_d(490)$. Within the BSTS framework, the control-zone SST and SSS serve, together with the control-zone $K_d(490)$, as exogenous covariates that absorb basin-scale climate variability (ENSO, the IOD, the monsoon reversal, and the Madden--Julian Oscillation), so that any residual post-intervention divergence between the observed and counterfactual impact-zone $K_d(490)$ can be attributed to local forcing.

\section{Methods}

\subsection{LULC Change Analysis}
\label{sec:lulc}

Annual LULC maps over the impact zone were extracted from the Esri 10~m product~\cite{karra2021} for eight consecutive years ($Y = \{2017, 2018, \ldots, 2024\}$), yielding $2226 \times 2226$ pixel rasters at 10~m resolution from the Sentinel-2 constellation~\cite{drusch2012}. Water pixels (class code~1) and cloud-masked pixels (class code~10) were excluded, retaining $J = 6$ terrestrial classes: trees, flooded vegetation, crops, built area, bare ground, and rangeland. For each pair of consecutive annual maps $(Y_t, Y_{t+1})$, a pixel-level cross-tabulation matrix $\mathbf{C}^{(t)} = [C_{ij}^{(t)}]$ of dimension $J \times J$ was constructed, where $C_{ij}^{(t)}$ denotes the number of pixels classified as category $i$ at time $Y_t$ and category $j$ at time $Y_{t+1}$, with $i, j \in \{1, 2, \ldots, J\}$. Seven such matrices ($t = 1, 2, \ldots, 7$) span the study period. The diagonal element $C_{ii}^{(t)}$ records the number of pixels that persisted in category $i$, while the off-diagonal sum
\begin{equation}
\Delta^{(t)} = \sum_{i=1}^{J}\sum_{\substack{j=1 \\ j \neq i}}^{J} C_{ij}^{(t)}
\label{eq:delta}
\end{equation}
gives the total changed pixels. An aggregated matrix $\mathbf{A} = [A_{ij}]$, where $A_{ij} = \sum_{t=1}^{7} C_{ij}^{(t)}$, was computed by summing elementwise across all seven intervals. All computations, intensity analysis, and statistical tests were implemented in Python using NumPy~\cite{harris2020}, SciPy~\cite{virtanen2020}, and netCDF4, with figures rendered via Matplotlib~\cite{hunter2007}.

The three-level intensity analysis framework of Aldwaik and Pontius~\cite{aldwaik2012} was applied to determine whether observed LULC changes were distributed uniformly across intervals, categories, and specific transitions, or whether systematic deviations indicative of targeted land conversion could be identified. At the first level (interval intensity), the annual change intensity $S_t$ for each interval is
\begin{equation}
S_t = \frac{\Delta^{(t)}}{N^{(t)}} \times \frac{100}{d_t},
\label{eq:St}
\end{equation}
where $\Delta^{(t)}$ is from equation~(\ref{eq:delta}), $N^{(t)} = \sum_{i}\sum_{j} C_{ij}^{(t)}$ is the total valid land pixels during interval $t$, and $d_t = Y_{t+1} - Y_t$ is the interval duration in years. The uniform threshold is
\begin{equation}
U_{\mathrm{int}} = \frac{\sum_{t=1}^{T} \Delta^{(t)}}{\sum_{t=1}^{T} N^{(t)}} \times \frac{100}{Y_T - Y_1},
\label{eq:Uint}
\end{equation}
where $T = 7$ and $Y_1 = 2017$, $Y_T = 2024$. An interval is \emph{active} if $S_t > U_{\mathrm{int}}$ and \emph{dormant} otherwise. Departure from uniformity was tested via a Pearson chi-square statistic $\chi^2 = \sum_{t} (O_t - E_t)^2 / E_t$, where $O_t = \Delta^{(t)}$ and $E_t = (N^{(t)} d_t / \sum_{t'} N^{(t')} d_{t'}) \sum_{t'} \Delta^{(t')}$.

At the second level (category intensity), let $A_{i+} = \sum_{j} A_{ij}$ denote the row marginal (start-period size of category $i$) and $A_{+j} = \sum_{i} A_{ij}$ the column marginal (end-period size of category $j$). The gain intensity $G_j$ and loss intensity $L_i$ are
\begin{equation}
G_j = \frac{A_{+j} - A_{jj}}{A_{+j}} \times 100, \qquad
L_i = \frac{A_{i+} - A_{ii}}{A_{i+}} \times 100.
\label{eq:GjLi}
\end{equation}
The uniform category threshold is
\begin{equation}
U_{\mathrm{cat}} = \frac{\sum_{i}\sum_{j \neq i} A_{ij}}{\sum_{i}\sum_{j} A_{ij}} \times 100.
\label{eq:Ucat}
\end{equation}
A category with $G_j$ (or $L_i$) exceeding $U_{\mathrm{cat}}$ is classified as active; otherwise it is dormant. Chi-square tests were applied separately for gains and losses.

At the third level (transition intensity), the transition intensity $R_{ij}$ for the gain of category $j$ from source $i$ ($i \neq j$) and the uniform gain threshold $W_j$ are
\begin{equation}
R_{ij} = \frac{A_{ij}}{A_{i+}} \times 100, \qquad
W_j = \frac{\mathrm{Gain}_j}{\sum_{i \neq j} A_{i+}} \times 100,
\label{eq:RijWj}
\end{equation}
where $\mathrm{Gain}_j = A_{+j} - A_{jj}$. A source $i$ is \emph{targeted} if $R_{ij} > W_j$ and \emph{avoided} otherwise. For the loss of category $j$ to target $k$ ($k \neq j$):
\begin{equation}
Q_{jk} = \frac{A_{jk}}{A_{+k}} \times 100, \qquad
V_j = \frac{\mathrm{Loss}_j}{\sum_{k \neq j} A_{+k}} \times 100,
\label{eq:QjkVj}
\end{equation}
where $\mathrm{Loss}_j = A_{j+} - A_{jj}$. Chi-square tests at this level used expected counts proportional to $A_{i+}$ (gain) or $A_{+k}$ (loss). Practical significance was quantified by Cohen's $h$~\cite{cohen1988}:
\begin{equation}
h = 2\arcsin\!\sqrt{p_{\mathrm{obs}}} - 2\arcsin\!\sqrt{p_{\mathrm{exp}}},
\label{eq:cohenh}
\end{equation}
where $p_{\mathrm{obs}}$ and $p_{\mathrm{exp}}$ are the observed and expected transition proportions from equations~(\ref{eq:RijWj})--(\ref{eq:QjkVj}). Following convention, $|h| \geq 0.8$ is large, $\geq 0.5$ medium, $\geq 0.2$ small, and $< 0.2$ negligible.

The nature of landscape change was decomposed following the quantity--exchange--shift (QES) framework of Pontius~\cite{pontius2019}. For each category $j$:
\begin{equation}
\mathcal{Q}_j = |\mathrm{Gain}_j - \mathrm{Loss}_j|, \qquad
\mathcal{E}_j = 2\sum_{i \neq j} \min(A_{ij}, A_{ji}), \qquad
\mathcal{S}_j = 2\min(\mathrm{Gain}_j, \mathrm{Loss}_j) - \mathcal{E}_j,
\label{eq:QES}
\end{equation}
where $\mathcal{Q}_j$, $\mathcal{E}_j$, and $\mathcal{S}_j$ denote the quantity, exchange, and shift components. Landscape-level totals were obtained by halving the category sums to avoid double-counting: $\mathcal{Q} = \tfrac{1}{2}\sum_{j} \mathcal{Q}_j$, $\mathcal{E} = \tfrac{1}{2}\sum_{j} \mathcal{E}_j$, $\mathcal{S} = \tfrac{1}{2}\sum_{j} \mathcal{S}_j$, with $\mathcal{T} = \mathcal{Q} + \mathcal{E} + \mathcal{S}$.

Temporal stationarity of the transition probability structure was assessed via a likelihood-ratio $G$-test~\cite{anderson1957}. Under $H_0$ that the row-conditional transition probability matrix $\mathbf{P} = [p_{ij}]$ is constant across all $T$ intervals:
\begin{equation}
G = 2\sum_{t=1}^{T}\sum_{i=1}^{J}\sum_{j=1}^{J} n_{ij}^{(t)} \ln\!\left(\frac{\hat{p}_{ij}^{(t)}}{\hat{p}_{ij}}\right),
\label{eq:Gtest}
\end{equation}
where $n_{ij}^{(t)} = C_{ij}^{(t)}$, $\hat{p}_{ij}^{(t)} = C_{ij}^{(t)} / C_{i+}^{(t)}$, and $\hat{p}_{ij} = A_{ij} / A_{i+}$. Terms with $n_{ij}^{(t)} = 0$ were omitted. Under $H_0$, $G \sim \chi^2$ with $(T-1) \times J \times (J-1)$ degrees of freedom. Per-class stationarity was evaluated by restricting to a single source row $i$, yielding $G_i$ with $(T-1)(J-1)$ degrees of freedom.

All confidence intervals on proportions were computed using the Wilson score method~\cite{wilson1927}. For $k$ successes out of $n$ trials at confidence level $1 - \alpha$, let $\hat{p} = k/n$ and $z_{\alpha/2}$ denote the upper $\alpha/2$ quantile of the standard normal. The Wilson interval is
\begin{equation}
\tilde{p} \pm \frac{z_{\alpha/2}}{1 + z_{\alpha/2}^2/n}\,
\sqrt{\frac{\hat{p}(1-\hat{p})}{n} + \frac{z_{\alpha/2}^2}{4n^2}},
\label{eq:wilson}
\end{equation}
where $\tilde{p} = (\hat{p} + z_{\alpha/2}^2 / 2n) / (1 + z_{\alpha/2}^2 / n)$. All tests in this subsection used $\alpha = 0.05$.

\subsection{Exploratory Analysis of Oceanographic Time Series}
\label{sec:exploratory}

The three oceanographic variables, $K_d(490)$ (in units of $\times 10^{-2}$~m$^{-1}$), SST ($^{\circ}$C), and SSS (PSU), were extracted as area-weighted spatial means over each domain for every month. Let $\phi_k$ denote the latitude of the $k$-th grid cell in a domain with $K$ valid pixels. The spatial mean of a generic variable $X$ is
\begin{equation}
\bar{X}(t) = \frac{\sum_{k=1}^{K} w_k \, X_k(t)}{\sum_{k=1}^{K} w_k}, \qquad w_k = \cos\phi_k,
\label{eq:cosweight}
\end{equation}
where $X_k(t)$ is the value at pixel $k$ and month $t$. Isolated temporal gaps were filled by linear interpolation prior to spatial averaging. Extraction and weighting used xarray~\cite{hoyer2017} and pandas~\cite{mckinney2010}; subsequent statistical computations used NumPy~\cite{harris2020} and SciPy~\cite{virtanen2020}.

The monthly time series $\{\bar{X}(t)\}_{t=1}^{n}$ ($n = 324$) were partitioned into four policy-aligned epochs: a pre-smelter baseline (before April 2015), initial IMIP operations (April 2015 to December 2019), post-export-ban hyper-expansion (January 2020 onward), and the full record. For each epoch and zone, robust location and scale measures were computed. The sample median $\tilde{X}$ served as the primary central tendency measure. Dispersion was characterized by the interquartile range ($\mathrm{IQR} = Q_3 - Q_1$) and the median absolute deviation
\begin{equation}
\mathrm{MAD} = \mathrm{med}_{i}\,\bigl| X_i - \tilde{X} \bigr|.
\label{eq:mad}
\end{equation}
Distributional shape was summarized by the sample skewness
\begin{equation}
g_1 = \frac{n}{(n-1)(n-2)} \sum_{i=1}^{n} \left(\frac{X_i - \bar{X}}{s}\right)^{\!3}
\label{eq:skewness}
\end{equation}
and the excess kurtosis
\begin{equation}
g_2 = \frac{n(n+1)}{(n-1)(n-2)(n-3)} \sum_{i=1}^{n} \left(\frac{X_i - \bar{X}}{s}\right)^{\!4} - \frac{3(n-1)^2}{(n-2)(n-3)},
\label{eq:kurtosis}
\end{equation}
where $\bar{X}$ and $s$ are the sample mean and standard deviation.

The annual cycle of each variable was characterized by grouping the full record by calendar month $m \in \{1, \ldots, 12\}$. For each month, the median, IQR, and a 95\% confidence interval for the median were computed via the nonparametric bootstrap~\cite{efron1993} with $B = 10{,}000$ resamples and a fixed random seed.

Monotonic trends in the $K_d(490)$ series were estimated using the Theil--Sen estimator~\cite{theil1950,sen1968}:
\begin{equation}
\hat{\beta}_{\mathrm{TS}} = \mathrm{med}_{i < j}\,\frac{X_j - X_i}{t_j - t_i},
\label{eq:theilsen}
\end{equation}
with intercept $\hat{\alpha}_{\mathrm{TS}} = \tilde{X} - \hat{\beta}_{\mathrm{TS}}\,\tilde{t}$. Statistical significance was assessed by Kendall's $\tau$~\cite{kendall1975}:
\begin{equation}
\tau = \frac{n_c - n_d}{\tfrac{1}{2}\,n(n-1)},
\label{eq:kendall}
\end{equation}
where $n_c$ and $n_d$ are the concordant and discordant pair counts, respectively. Two-sided $p$-values were computed from the asymptotic normal distribution. The analysis was applied separately to each zone and each epoch.

\subsection{Structural Break Detection}
\label{sec:changepoint}

Abrupt regime shifts in $K_d(490)$ were assessed through a multi-algorithm consensus changepoint framework applied independently to each zone. Let $\{x_t\}_{t=1}^{n}$ denote the monthly area-weighted $K_d(490)$ for a given domain. The objective is to partition the record into $k+1$ contiguous segments delimited by changepoints $\boldsymbol{\tau} = \{\tau_1, \ldots, \tau_k\}$ such that the signal is approximately piecewise stationary. A radial basis function (RBF) kernel cost was adopted as the segment homogeneity measure, making no parametric assumption on the within-segment distribution~\cite{truong2020,celisse2018}. For a candidate segment $(a, b]$:
\begin{equation}
\mathcal{C}(x_{a+1:b}) = n_{ab} - \frac{1}{n_{ab}} \sum_{i=a+1}^{b}\sum_{j=a+1}^{b} \exp\!\bigl(-\gamma\,(x_i - x_j)^2\bigr),
\label{eq:rbfcost}
\end{equation}
where $n_{ab} = b - a$ and $\gamma > 0$ is the kernel bandwidth (set to the inverse median of squared pairwise distances).

Three algorithms solved the penalized segmentation problem $\min_{\boldsymbol{\tau}} V(\boldsymbol{\tau}) + f(k)$: (1)~Pruned Exact Linear Time (PELT)~\cite{killick2012} with a linear penalty $f(k) = \beta k$; (2)~Binary Segmentation (BinSeg)~\cite{scott1974}; and (3)~window-based detection~\cite{truong2020}. All were implemented via the \texttt{ruptures} library~\cite{truong2020} with minimum segment length~3.

The penalty $\beta$ for PELT was selected via a sensitivity sweep over $\beta \in [1, 50]$, with the optimal $\beta^*$ identified by the discrete elbow criterion:
\begin{equation}
\beta^* = \beta_{\,\arg\max_\ell\, |\Delta^2 \hat{k}(\beta_\ell)|},
\label{eq:elbow}
\end{equation}
where $\Delta^2 \hat{k}(\beta_\ell)$ is the second finite difference of the smoothed changepoint count. The number of changepoints was independently selected by the Bayesian Information Criterion (BIC)~\cite{schwarz1978}:
\begin{equation}
\mathrm{BIC}(k) = n \ln\!\left(\frac{\mathrm{RSS}(k)}{n}\right) + p_k \ln n,
\label{eq:bic}
\end{equation}
where $\mathrm{RSS}(k)$ is the residual sum of squares and $p_k = 2(k+1)$. The BIC-optimal order is $k^* = \arg\min_k \mathrm{BIC}(k)$.

A consensus rule required agreement of at least two of three algorithms within a $\pm 5$-month tolerance window. The significance of each consensus breakpoint was assessed by a nonparametric permutation test~\cite{efron1993}. For a breakpoint at index $\tau$:
\begin{equation}
D_{\mathrm{obs}} = \bigl|\bar{x}_{1:\tau} - \bar{x}_{\tau+1:n}\bigr|.
\label{eq:Dobs}
\end{equation}
A reference distribution was constructed from $B_{\mathrm{perm}} = 5{,}000$ random permutations, and the $p$-value estimated as
\begin{equation}
\hat{p}_{\mathrm{perm}} = \frac{1}{B_{\mathrm{perm}}} \sum_{b=1}^{B_{\mathrm{perm}}} \mathbf{1}\!\bigl(D^{*(b)} \geq D_{\mathrm{obs}}\bigr).
\label{eq:pperm}
\end{equation}

For each pair of adjacent regimes, four complementary tests quantified the distributional shift: Welch's $t$-test~\cite{welch1947}:
\begin{equation}
t_W = \frac{\bar{x}_1 - \bar{x}_2}{\sqrt{s_1^2/n_1 + s_2^2/n_2}},
\label{eq:welch}
\end{equation}
with Welch--Satterthwaite degrees of freedom; the Mann--Whitney $U$ test~\cite{mann1947}; Levene's test~\cite{levene1960}; and the two-sample Kolmogorov--Smirnov (KS) test~\cite{massey1951}. The nonparametric effect size was measured by Cliff's $\delta$~\cite{cliff1993}:
\begin{equation}
\delta_C = \frac{1}{n_1 n_2}\sum_{i=1}^{n_1}\sum_{j=1}^{n_2} \mathrm{sgn}(x_{1,i} - x_{2,j}),
\label{eq:cliff}
\end{equation}
with thresholds of Romano et al.~\cite{romano2006}: $|\delta_C| < 0.147$ negligible, $< 0.33$ small, $< 0.474$ medium, and $\geq 0.474$ large. The effective sample size in each regime was adjusted for temporal autocorrelation following Bretherton et al.~\cite{bretherton1999}:
\begin{equation}
n_{\mathrm{eff}} = n_r \,\frac{1 - \rho_1}{1 + \rho_1},
\label{eq:neff}
\end{equation}
where $\rho_1$ is the lag-1 autocorrelation.

Spatial specificity was evaluated by a difference-in-differences (DiD) estimator~\cite{angrist2009}:
\begin{equation}
\widehat{\mathrm{DiD}} = \bigl(\bar{x}^{I}_{\mathrm{post}} - \bar{x}^{I}_{\mathrm{pre}}\bigr) - \bigl(\bar{x}^{C}_{\mathrm{post}} - \bar{x}^{C}_{\mathrm{pre}}\bigr),
\label{eq:did}
\end{equation}
where superscripts $I$ and $C$ denote impact and control zones. Under the identifying assumption that the control zone captures all shared basin-scale forcing, a nonzero $\widehat{\mathrm{DiD}}$ is attributable to local perturbation.

\subsection{BSTS Causal Impact Estimation}
\label{sec:bsts}

The causal effect of local perturbation on impact-zone $K_d(490)$ was quantified within the BSTS framework of Brodersen et al.~\cite{brodersen2015}. The intervention date was set to $t^* = \hat{\tau}_1$, the earliest consensus changepoint, partitioning the record into a pre-intervention period $\mathcal{T}_{\mathrm{pre}} = \{1, \ldots, t^* - 1\}$ and a post-intervention period $\mathcal{T}_{\mathrm{post}} = \{t^*, \ldots, n\}$. Denoting the impact-zone monthly mean $K_d(490)$ by $y_t$ and collecting the three control-zone covariates into $\mathbf{z}_t = (z_{1,t},\, z_{2,t},\, z_{3,t})^\top$ (control-zone $K_d(490)$, SST, and SSS), the observation equation is
\begin{equation}
y_t = \mu_t + \gamma_t + \boldsymbol{\beta}^\top \mathbf{z}_t + \varepsilon_t, \qquad \varepsilon_t \sim \mathcal{N}(0, \sigma_\varepsilon^2),
\label{eq:bsts_obs}
\end{equation}
where $\mu_t$ is a local linear trend, $\gamma_t$ is a stochastic seasonal component with period $s = 12$ months, $\boldsymbol{\beta}$ is a static regression coefficient vector, and $\varepsilon_t$ is Gaussian observation noise. The trend evolves as
\begin{align}
\mu_{t+1} &= \mu_t + \nu_t + \eta_t^{\mu}, &\quad \eta_t^{\mu} &\sim \mathcal{N}(0, \sigma_\mu^2), \label{eq:bsts_level}\\
\nu_{t+1} &= \nu_t + \eta_t^{\nu}, &\quad \eta_t^{\nu} &\sim \mathcal{N}(0, \sigma_\nu^2), \label{eq:bsts_slope}
\end{align}
where $\nu_t$ is the stochastic slope and $\eta_t^{\mu}$, $\eta_t^{\nu}$ are independent Gaussian disturbances~\cite{harvey1989,durbin2012}. The seasonal component satisfies
\begin{equation}
\sum_{j=0}^{s-1} \gamma_{t-j} = \eta_t^{\gamma}, \qquad \eta_t^{\gamma} \sim \mathcal{N}(0, \sigma_\gamma^2),
\label{eq:bsts_seasonal}
\end{equation}
permitting gradual evolution of the annual cycle. The hyperparameters were estimated by maximum likelihood via L-BFGS-B (maximum $1{,}000$ iterations), with the log-likelihood evaluated by Kalman filter recursions~\cite{durbin2012}. The first 13~months of filtered output were discarded as burn-in. The model was implemented using the \texttt{UnobservedComponents} class of \texttt{statsmodels}~\cite{seabold2010}.

The model was fitted on $\mathcal{T}_{\mathrm{pre}}$ and projected forward over $\mathcal{T}_{\mathrm{post}}$ using observed $\mathbf{z}_t$ to produce a counterfactual. Let $\hat{y}_t$ denote the counterfactual prediction and $[\hat{y}_t^{L},\, \hat{y}_t^{U}]$ its 95\% prediction interval. The pointwise and cumulative causal effects are
\begin{equation}
\delta_t = y_t - \hat{y}_t, \qquad
\Delta_T = \sum_{t=t^*}^{T} \delta_t.
\label{eq:causal_effects}
\end{equation}
The average effect is $\bar{\delta} = n_{\mathrm{post}}^{-1} \sum_{t \in \mathcal{T}_{\mathrm{post}}} \delta_t$, and the relative effect is $\bar{\delta}_{\mathrm{rel}} = \bar{\delta}\,/\,\bar{\hat{y}} \times 100\,\%$, where $\bar{\hat{y}}$ is the mean counterfactual. Significance was assessed by a two-sided $z$-test:
\begin{equation}
z = \frac{|\bar{\delta}|}{\mathrm{SE}}, \qquad \mathrm{SE} = \frac{\bar{\hat{y}}^{U} - \bar{\hat{y}}^{L}}{2 \times 1.96},
\label{eq:bsts_ztest}
\end{equation}
with $p = 2\,[1 - \Phi(z)]$~\cite{brodersen2015}.

Three robustness checks were performed. First, a placebo test assessed whether an effect of comparable magnitude could arise spuriously during $\mathcal{T}_{\mathrm{pre}}$. A total of $N_{\mathrm{plac}} = 40$ pseudo-intervention dates were drawn uniformly at random (seed~=~42) from $\{37, \ldots, |\mathcal{T}_{\mathrm{pre}}| - 12\}$, reserving 36~months for fitting and 12 for a pseudo-post period. For each pseudo-date $t_j^*$, the BSTS model was re-estimated on $\{1, \ldots, t_j^* - 1\}$ and $\bar{\delta}_j^{\mathrm{plac}}$ computed over the subsequent 12~months. The rank-based $p$-value is
\begin{equation}
\hat{p}_{\mathrm{rank}} = \frac{1}{N_{\mathrm{valid}}} \sum_{j=1}^{N_{\mathrm{valid}}} \mathbf{1}\!\bigl(|\bar{\delta}_j^{\mathrm{plac}}| \geq |\bar{\delta}|\bigr),
\label{eq:rankp}
\end{equation}
providing a distribution-free confirmation independent of Gaussian assumptions~\cite{brodersen2015}.

Second, a leave-one-out covariate sensitivity analysis re-estimated the model four times (once with all covariates and once with each covariate removed) to verify sign and significance stability. Third, pre-period residual diagnostics evaluated model adequacy via the Shapiro--Wilk test for normality~\cite{shapiro1965} and the Ljung--Box portmanteau statistic at $m = 12$ lags~\cite{ljung1978}:
\begin{equation}
Q_{\mathrm{LB}} = n_{\mathrm{pre}}(n_{\mathrm{pre}} + 2) \sum_{\ell=1}^{m} \frac{\hat{r}_\ell^{\,2}}{n_{\mathrm{pre}} - \ell},
\label{eq:ljungbox}
\end{equation}
where $\hat{r}_\ell$ is the sample autocorrelation at lag $\ell$, with $Q_{\mathrm{LB}} \sim \chi^2_m$ under $H_0$. Rejection signals model misspecification but does not invalidate $\hat{p}_{\mathrm{rank}}$, which is distribution-free.

The ecological relevance of the causal effect was expressed as a change in euphotic depth $Z_{\mathrm{eu}}$, the depth at which photosynthetically active radiation falls to 1\% of its surface value. Following Lee et al.~\cite{lee2007}:
\begin{equation}
Z_{\mathrm{eu}} \approx \frac{4.6}{K_d(490)}.
\label{eq:zeu}
\end{equation}
Counterfactual and observed euphotic depths were computed from $\bar{\hat{y}}$ and $\bar{y}_{\mathrm{post}}$ after unit conversion. The difference $\Delta Z_{\mathrm{eu}} = Z_{\mathrm{eu}}^{\mathrm{obs}} - Z_{\mathrm{eu}}^{\mathrm{cf}}$ quantifies photic zone change attributable to local perturbation.

\section{Results}

\begin{figure}[H]
\centering
\includegraphics[width=\textwidth]{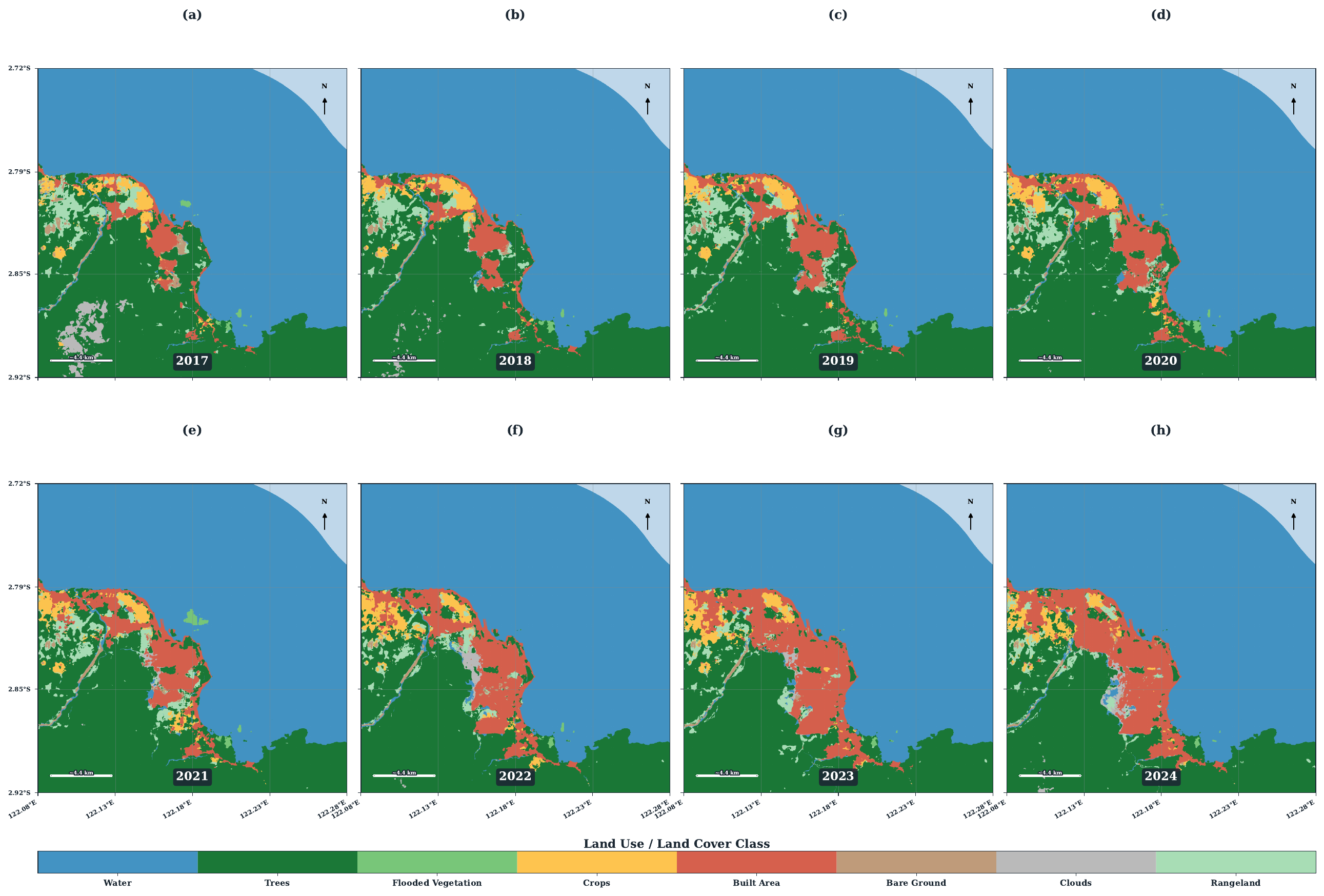}
\caption{Annual LULC maps of the IMIP impact zone derived from Sentinel-2 at 10~m resolution for (\textbf{a})~2017 through (\textbf{h})~2024. Eight classes are shown: water (blue), trees (dark green), flooded vegetation (light green), crops (orange), built area (red), bare ground (tan), clouds (gray), and rangeland (yellow-green).}
\label{fig:lulcmap}
\end{figure}

Annual LULC composites over the impact zone for 2017--2024 are shown in Figure~\ref{fig:lulcmap}. Built area expanded from $122{,}561$~pixels ($12.26$~km$^2$; 2.56\%) in 2017 to $461{,}789$~pixels ($46.18$~km$^2$; 9.65\%) in 2024, a net increase of $+7.09$ percentage points and a $3.8\times$ areal expansion. Tree cover fell from $1{,}548{,}511$~pixels ($154.85$~km$^2$; 32.37\%) to $1{,}307{,}457$~pixels ($130.75$~km$^2$; 27.33\%), a loss of $-5.04$ percentage points. Rangeland decreased from 2.35\% to 1.56\%, bare ground from 0.72\% to 0.30\%, crops fluctuated between 1.44\% and 1.98\%, and flooded vegetation remained below 0.31\%.

\begin{figure}[H]
\centering
\includegraphics[width=\textwidth]{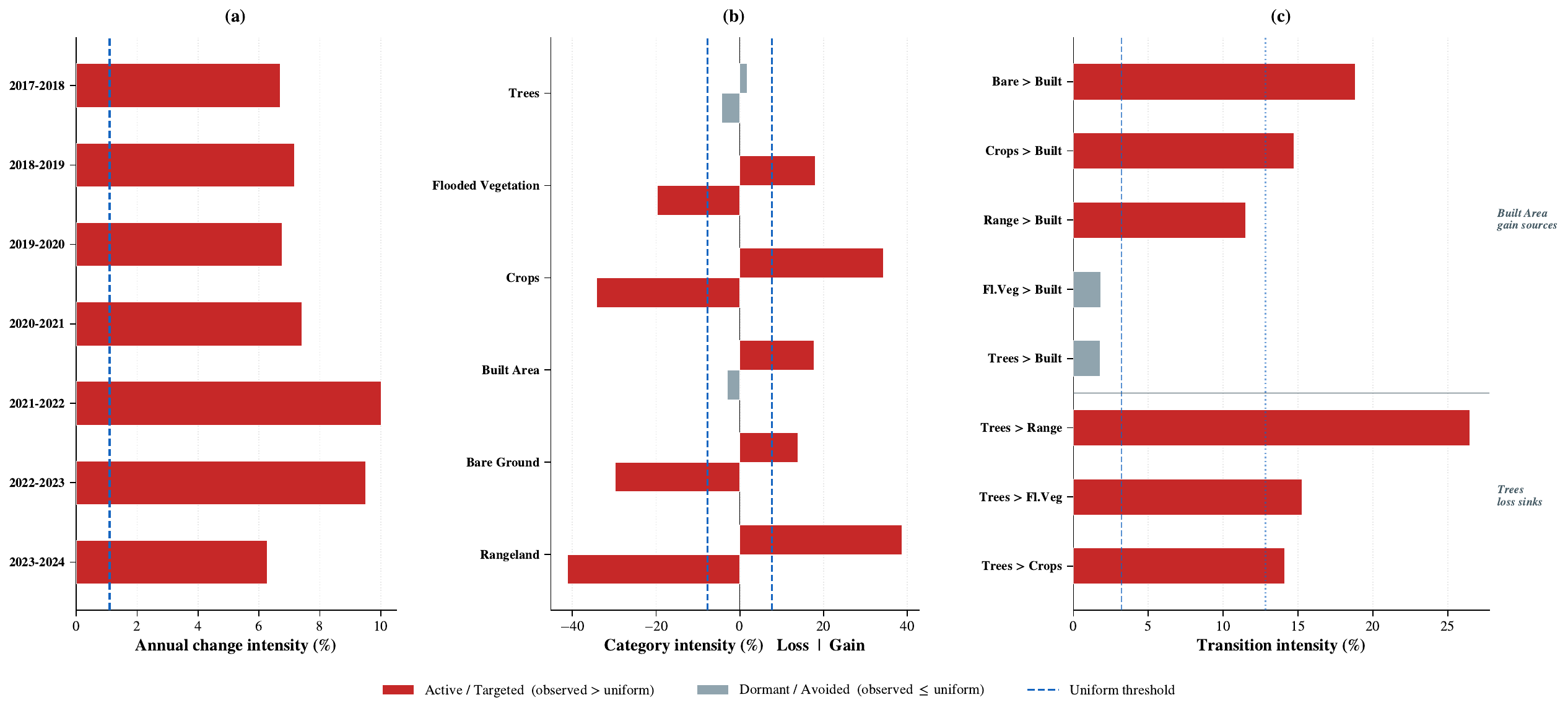}
\caption{Three-level intensity analysis of LULC transitions in the IMIP impact zone, 2017--2024. (\textbf{a})~Interval-level: annual change intensity $S_t$ for each year-pair; the dashed line marks $U_{\mathrm{int}}$. (\textbf{b})~Category-level: gain (right) and loss (left) intensities; dashed lines mark $U_{\mathrm{cat}}$. (\textbf{c})~Transition-level: selected intensities for built area gain sources and tree cover loss sinks; dashed lines mark the uniform thresholds. Red bars denote active/targeted transitions; gray bars denote dormant/avoided transitions.}
\label{fig:intensity}
\end{figure}

The interval-level intensity analysis (Figure~\ref{fig:intensity}a) classified all seven year-pairs as active, with $S_t$ exceeding $U_{\mathrm{int}} = 1.10\%$\,yr$^{-1}$ in every interval ($\chi^2 = 32{,}221$, $\mathrm{df} = 6$, $p < 0.001$). $S_t$ ranged from 6.27\% (2023--2024) to 10.01\% (2021--2022). Total changed area was $1{,}038{,}269$~pixels ($103.83$~km$^{2}$).

At the category level (Figure~\ref{fig:intensity}b), built area exhibited active gain ($G_j = 17.78\%$, exceeding $U_{\mathrm{cat}} = 7.68\%$) and dormant loss ($L_j = 2.95\%$), yielding a net gain of $+32.28$~km$^2$. Tree cover showed dormant gain (1.81\%) and dormant loss (4.39\%), with a net loss of $-26.98$~km$^2$. Rangeland registered active gain (38.75\%) and active loss (41.15\%), indicating high turnover. Crops exhibited active gain (34.31\%) and active loss (34.21\%) with negligible net change. Bare ground showed active gain (13.93\%) and active loss (29.70\%).

The transition-level analysis (Figure~\ref{fig:intensity}c) identified bare ground, crops, and rangeland as targeted sources for built area gain. Among sources ($W_j = 3.20\%$), bare ground yielded the highest $R_{ij}$ ($18.81\%$, $5.9\times$, $h = +0.537$, medium), followed by crops ($14.75\%$, $4.6\times$, $h = +0.428$, small) and rangeland ($11.52\%$, $3.6\times$, $h = +0.333$, small). Trees ($R_{ij} = 1.79\%$, $h = -0.092$) were avoided as a direct source. Among tree cover loss sinks, the highest intensities were trees to rangeland ($Q_{jk} = 26.47\%$, targeted), trees to flooded vegetation (15.27\%, targeted), and trees to crops (14.11\%, targeted), while trees to built area was avoided ($Q_{jk} = 8.68\%$ versus $V_j = 12.83\%$). The QES decomposition yielded quantity~$= 31.2\%$, exchange~$= 55.0\%$, and shift~$= 13.9\%$. The Markov stationarity test rejected $H_0$ ($G = 240{,}370$, $\mathrm{df} = 180$, $p < 0.001$); all six classes individually rejected stationarity.

\begin{figure}[H]
\centering
\includegraphics[width=0.7\textwidth]{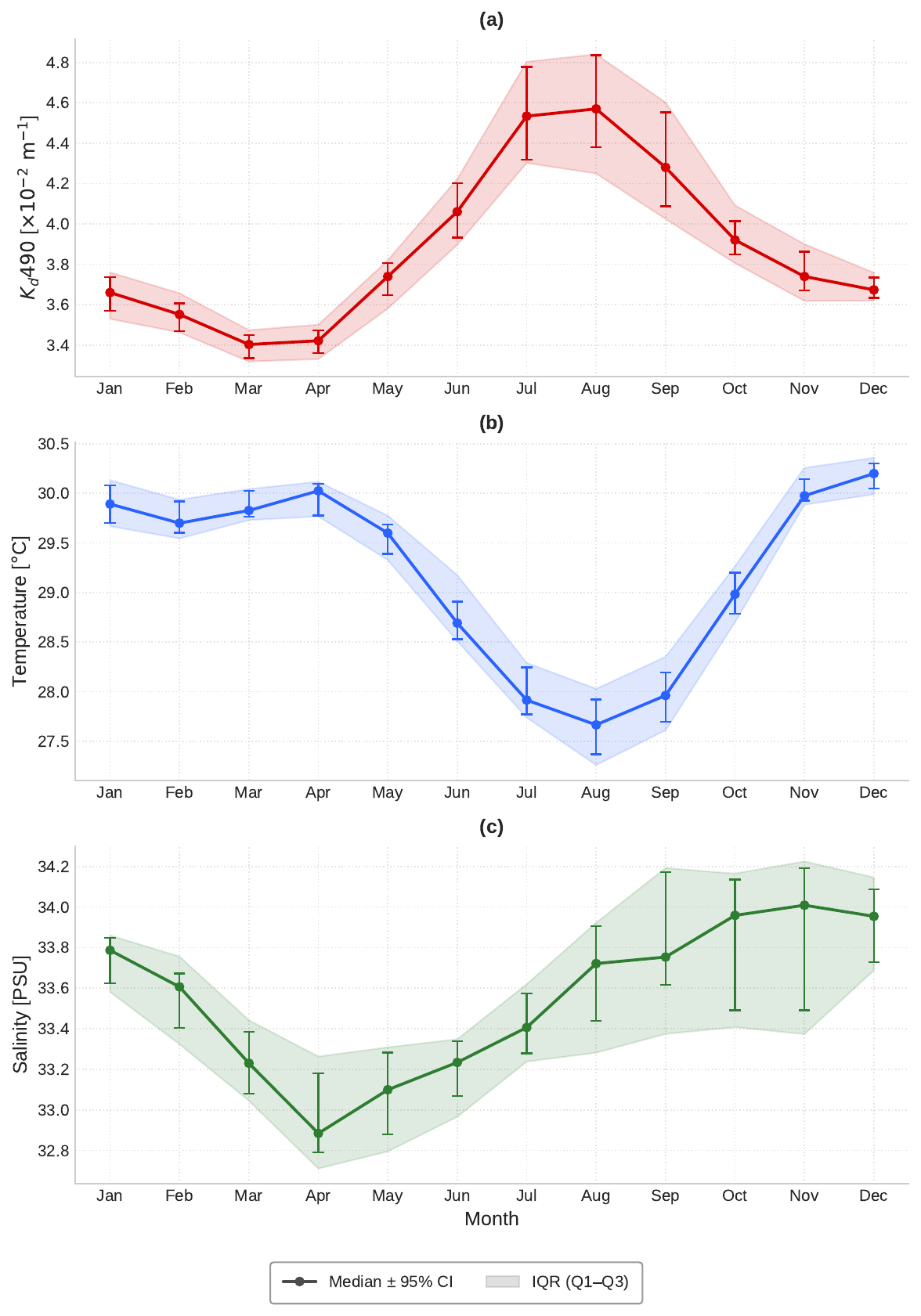}
\caption{Monthly climatology of the entire study area (1998--2024) for (\textbf{a})~$K_d(490)$ ($\times 10^{-2}$~m$^{-1}$), (\textbf{b})~SST ($^{\circ}$C), and (\textbf{c})~SSS (PSU). Points and error bars show the median and 95\% bootstrap confidence interval ($B = 10{,}000$); shading denotes the IQR.}
\label{fig:climatology}
\end{figure}

The monthly climatology (Figure~\ref{fig:climatology}) shows a pronounced annual cycle. $K_d(490)$ reached its minimum in March--April (median $\approx 3.40$--$3.42 \times 10^{-2}$~m$^{-1}$) and peaked in July--August ($\approx 4.53$--$4.57$), with larger IQR during boreal summer ($\approx 0.50$--$0.59$) than during March--April ($\approx 0.16$--$0.17$). SST peaked in December ($\approx 30.20$\,$^{\circ}$C) and was lowest in August ($\approx 27.67$\,$^{\circ}$C). SSS was lowest in April ($\approx 32.88$~PSU) and highest in November ($\approx 34.01$~PSU).

\begin{figure}[H]
\centering
\includegraphics[width=\textwidth]{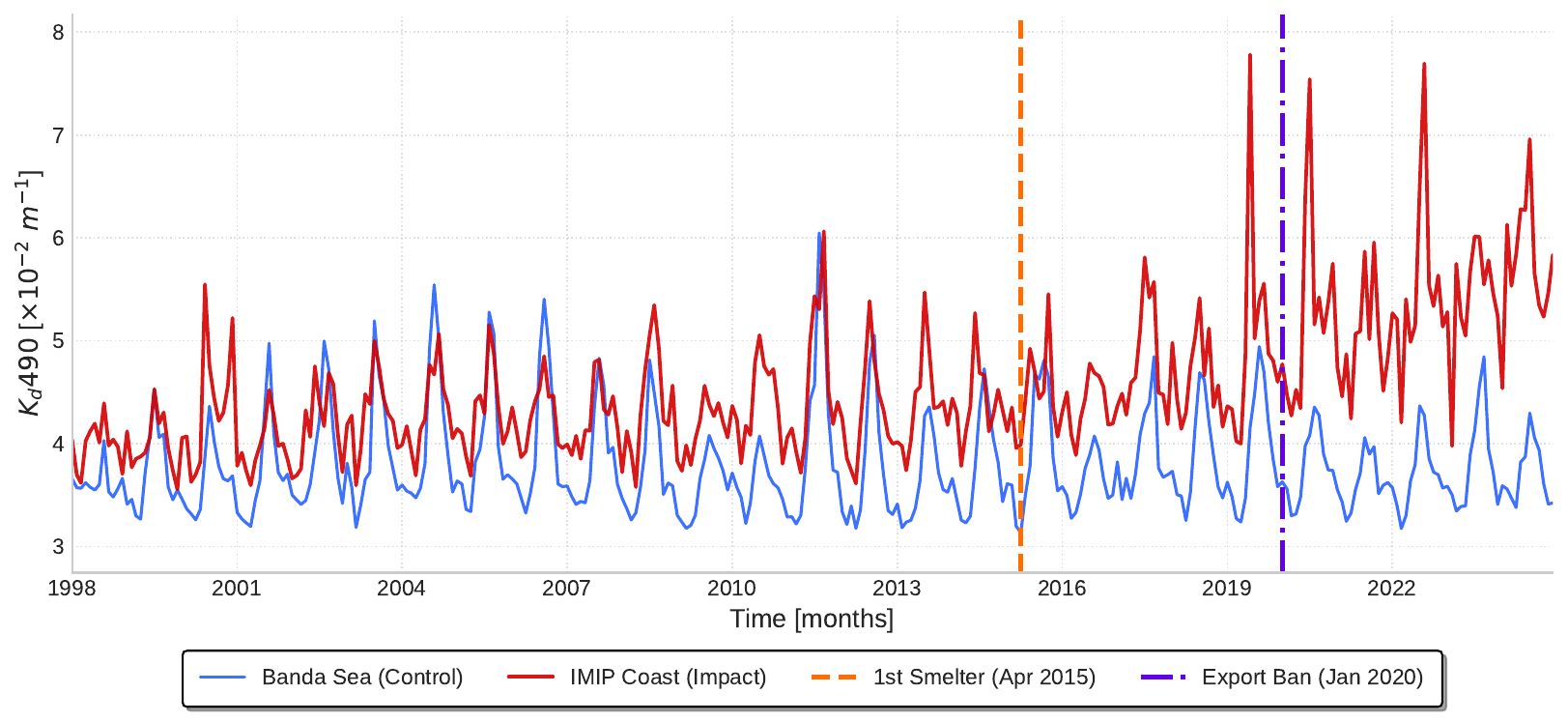}
\caption{Monthly $K_d(490)$ ($\times 10^{-2}$~m$^{-1}$) for the impact zone (red) and control zone (blue), January 1998 to December 2024. The orange dashed line marks first smelter commissioning (April 2015); the purple dash-dotted line marks the export ban (January 2020).}
\label{fig:rawts}
\end{figure}

The raw $K_d(490)$ time series (Figure~\ref{fig:rawts}) shows a full-record impact-zone median of $4.37$ (MAD~$= 0.37$), compared with $3.64$ (MAD~$= 0.23$) in the control zone. The impact-zone distribution is positively skewed ($g_1 = 1.52$, $g_2 = 3.48$). By epoch, impact-zone medians were $4.19$ (pre-smelter, $n = 207$), $4.51$ (initial operations, $n = 57$), and $5.31$ (post-ban, $n = 60$); control-zone medians were $3.62$, $3.70$, and $3.63$. Impact-zone MAD rose from $0.24$ to $0.27$ to $0.44$ across epochs, while control-zone MAD held steady at $0.23$--$0.24$.

The Theil--Sen estimator yielded a significant upward trend in the impact zone over the full record ($\hat{\beta}_{\mathrm{TS}} = +0.00384$, $\tau = 0.434$, $p < 0.001$), whereas the control zone showed no significant trend ($p = 0.736$). In the pre-smelter baseline, the impact zone showed a weaker but significant trend ($\hat{\beta}_{\mathrm{TS}} = +0.00171$, $\tau = 0.179$, $p < 0.001$); the control zone did not ($p = 0.621$). During initial operations, neither zone reached significance (impact: $p = 0.067$; control: $p = 0.670$). In the post-ban period, the impact-zone trend steepened ($\hat{\beta}_{\mathrm{TS}} = +0.01771$, $\tau = 0.316$, $p < 0.001$) while the control zone remained nonsignificant ($p = 0.702$).

\begin{figure}[H]
\centering
\includegraphics[width=\textwidth]{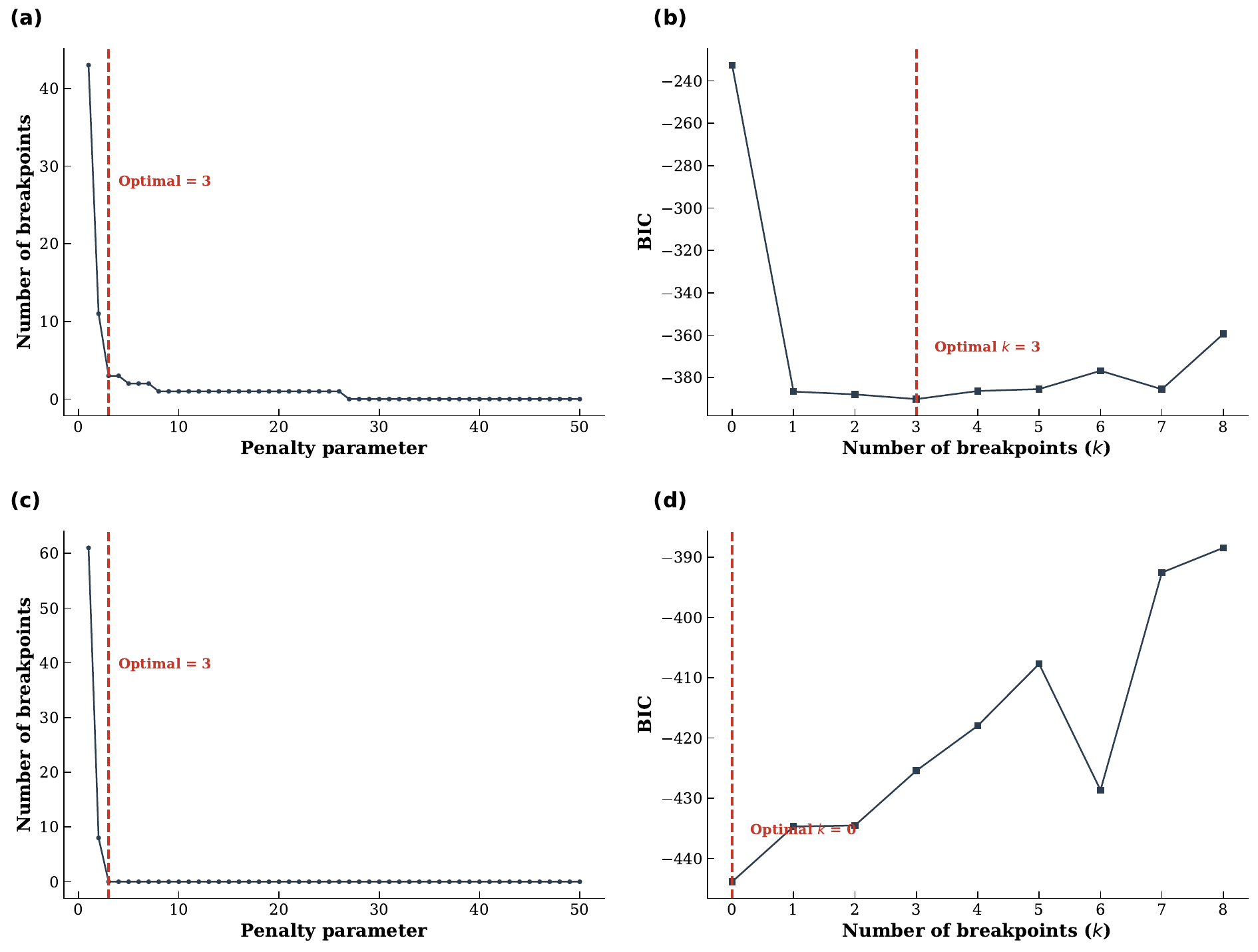}
\caption{Changepoint model selection for the impact zone (top) and control zone (bottom). (\textbf{a},~\textbf{c})~Number of breakpoints versus penalty parameter; the red dashed line marks the optimal penalty. (\textbf{b},~\textbf{d})~BIC versus $k$; the red dashed line marks the BIC-optimal $k$.}
\label{fig:penaltybic}
\end{figure}

\begin{figure}[H]
\centering
\includegraphics[width=\textwidth]{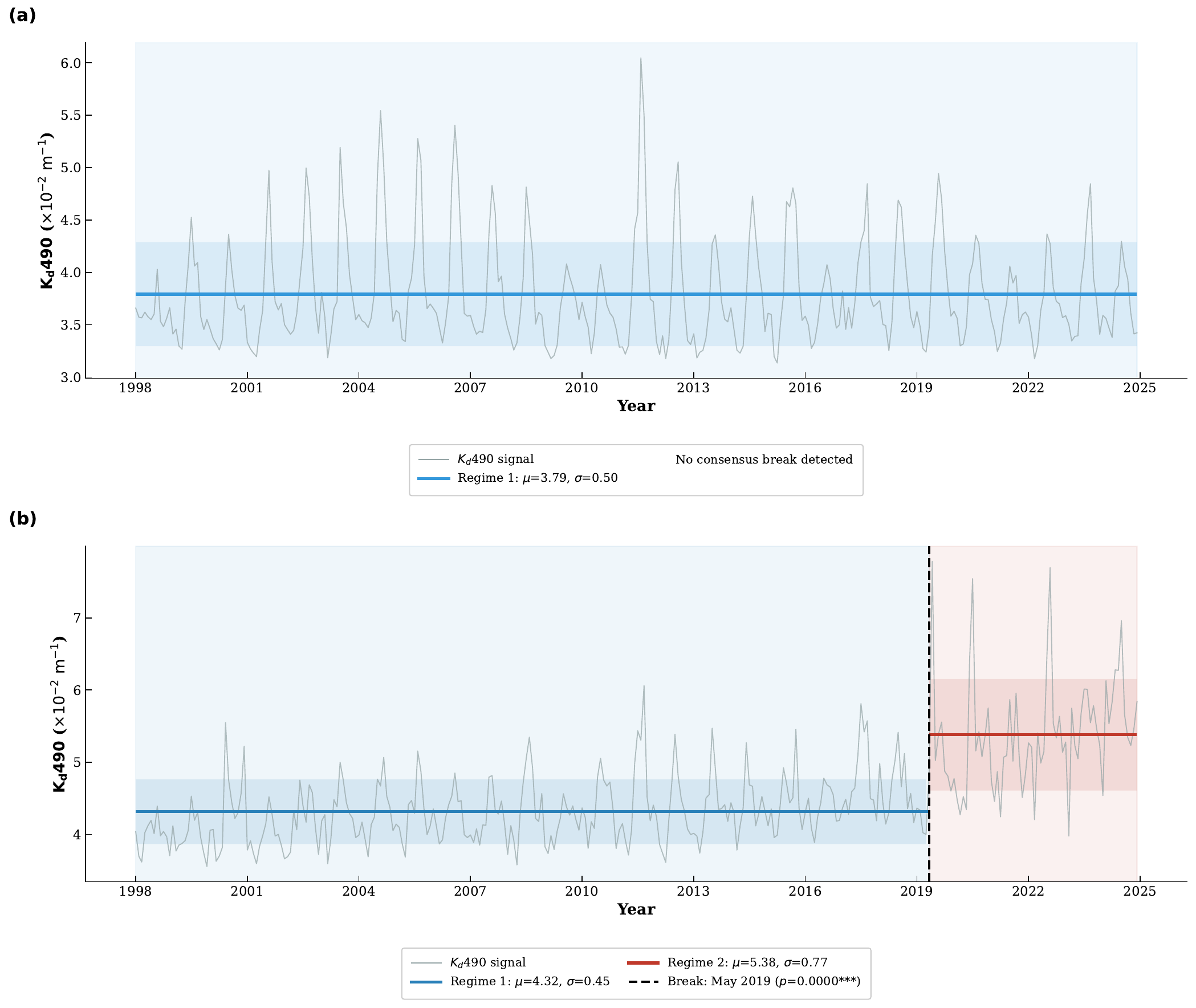}
\caption{Consensus regime segmentation of $K_d(490)$ for (\textbf{a})~the control zone and (\textbf{b})~the impact zone. Horizontal lines and shading indicate regime means $\pm 1\sigma$. In~(\textbf{a}), no consensus breakpoint was detected (single regime, $\mu = 3.79$, $\sigma = 0.50$). In~(\textbf{b}), a consensus breakpoint at May 2019 ($p < 0.001$) separates Regime~1 ($\mu = 4.32$, $\sigma = 0.45$) from Regime~2 ($\mu = 5.38$, $\sigma = 0.77$).}
\label{fig:regimes}
\end{figure}

Changepoint model selection for the impact zone (Figure~\ref{fig:penaltybic}a,~b) returned optimal $k = 3$ under both the penalty elbow and BIC criteria ($\mathrm{BIC} = -390.13$ at $k = 3$ versus $-232.58$ at $k = 0$). Individual algorithms identified: PELT, May 2019; BinSeg, April 2002, May 2019, July 2022; Window, June 2000, June 2019, June 2020. The consensus rule retained a single breakpoint at May 2019 (index~256), confirmed by permutation test ($B = 5{,}000$, $|\Delta\mu| = 1.066$, $p < 0.001$; Figure~\ref{fig:regimes}b). The record splits into Regime~1 ($n = 256$, $\mu = 4.32$, $\sigma = 0.45$, $\rho_1 = 0.595$) and Regime~2 ($n = 68$, $\mu = 5.38$, $\sigma = 0.77$, $\rho_1 = 0.275$). Pairwise comparisons yielded Cliff's $\delta = -0.806$ (large), Welch $t = -10.88$ ($p < 0.001$), Mann--Whitney $U = 1691.0$ ($p < 0.001$), KS $D = 0.692$ ($p < 0.001$), and Levene's $F = 19.50$ ($p < 0.001$).

For the control zone (Figure~\ref{fig:penaltybic}c,~d), BIC selected $k = 0$ ($\mathrm{BIC} = -443.88$) and the penalty curve collapsed to zero breakpoints at $\beta \geq 3$. No consensus breakpoint was detected (Figure~\ref{fig:regimes}a); the record remained a single regime ($n = 324$, $\mu = 3.79$, $\sigma = 0.50$, $\rho_1 = 0.734$). The DiD estimate was $\widehat{\mathrm{DiD}} = (+1.066) - (-0.021) = +1.086 \times 10^{-2}$~m$^{-1}$.

\begin{figure}[H]
\centering
\includegraphics[width=\textwidth]{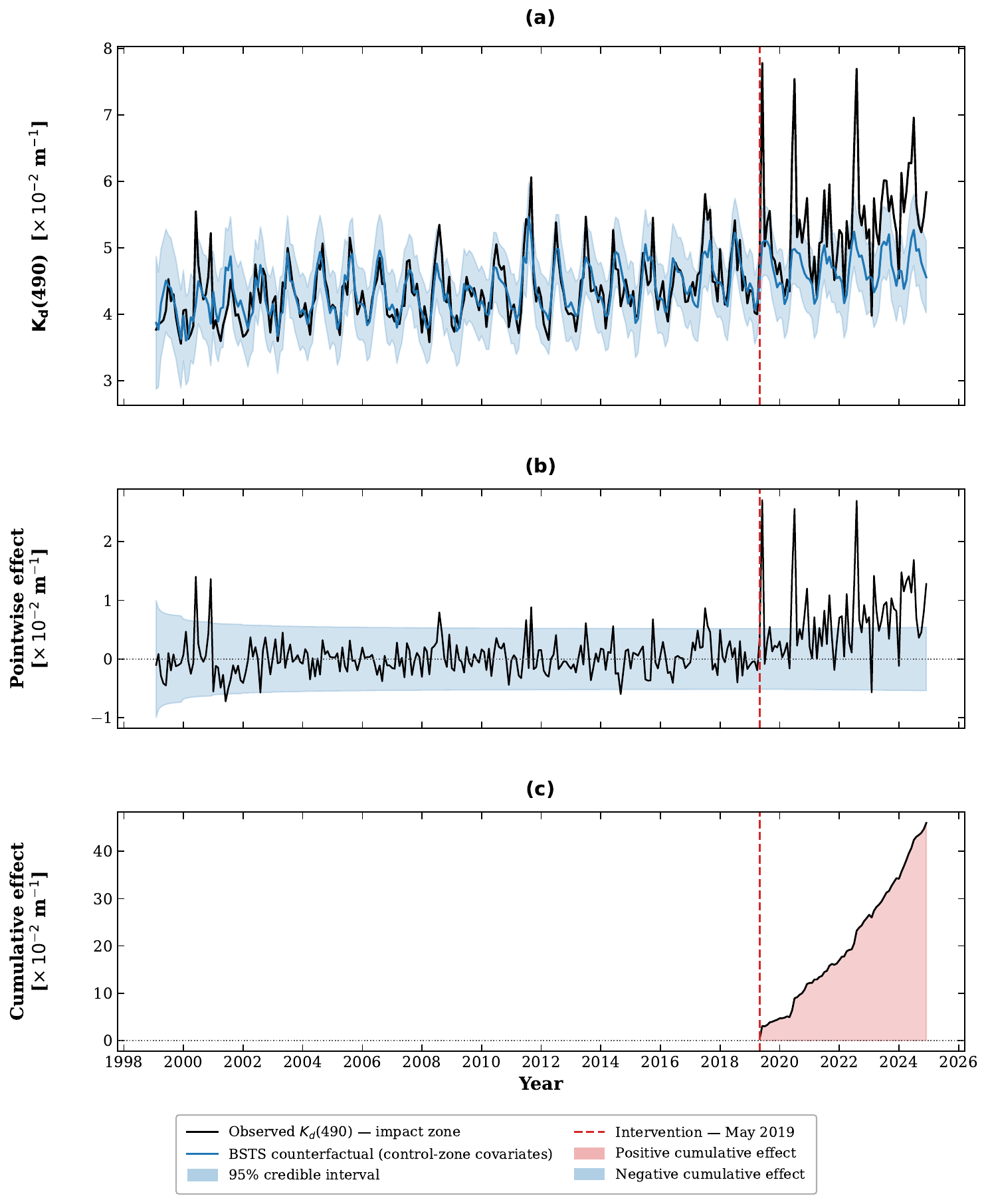}
\caption{BSTS causal impact analysis with intervention at May 2019. (\textbf{a})~Observed $K_d(490)$ (black) and counterfactual (blue) with 95\% credible interval. (\textbf{b})~Pointwise causal effect $\delta_t$ with 95\% interval. (\textbf{c})~Cumulative effect $\Delta_T$; red shading indicates positive accumulation, blue negative. The red dashed line marks the intervention.}
\label{fig:bsts}
\end{figure}

The BSTS model (Figure~\ref{fig:bsts}), fitted on the 256-month pre-period and projected over the 68-month post-period, yielded a post-period observed mean of $5.381 \times 10^{-2}$~m$^{-1}$ against a counterfactual mean of $4.704$ (95\% CI: $[4.179,\, 5.230]$). The average causal effect was $\bar{\delta} = +0.676$, corresponding to a relative effect of $+14.38\%$. The cumulative effect reached $\Delta_T = +46.00$ over 68 months and increased monotonically (Figure~\ref{fig:bsts}c). The $z$-test returned $z = 2.522$ ($p = 0.012$). Pointwise effects ranged from $-0.570$ (February 2023) to $+2.709$ (June 2019), with 63 of 68 post-intervention months positive (Figure~\ref{fig:bsts}b). Annual aggregate effects increased from $+0.547$ (2019, 8~months) to $+0.631$ (2020), $+0.361$ (2021), $+0.799$ (2022), $+0.704$ (2023), and $+0.973$ (2024).

The placebo test ($N_{\mathrm{plac}} = 40$, all converged) yielded a mean effect of $+0.041$ (SD~$= 0.111$; range $-0.178$ to $+0.325$); none exceeded $|\bar{\delta}|$, giving $\hat{p}_{\mathrm{rank}} = 0.000$. Leave-one-out sensitivity returned: all covariates, $\bar{\delta} = +0.676$ ($p = 0.012$); dropping control $K_d(490)$, $+0.714$ ($p = 0.018$); dropping control SST, $+0.676$ ($p = 0.013$); dropping control SSS, $+0.685$ ($p = 0.011$). All configurations produced the same positive sign and significance at $\alpha = 0.05$. Pre-period residual diagnostics ($n_{\mathrm{pre}} = 256$) rejected normality (Shapiro--Wilk $W = 0.697$, $p < 0.001$) and independence (Ljung--Box $p < 0.001$), with residual $g_1 = +3.86$ and $g_2 = +30.00$. Conversion to euphotic depth via equation~(\ref{eq:zeu}) yielded a counterfactual $Z_{\mathrm{eu}} = 97.8$~m and an observed $Z_{\mathrm{eu}} = 85.5$~m, giving $\Delta Z_{\mathrm{eu}} = -12.3$~m ($-12.6\%$).

\section{Discussion}

Four independent lines of evidence converge on the same conclusion: a multi-algorithm consensus breakpoint at May 2019 ($p < 0.001$), a BSTS posterior mean causal effect of $\bar{\delta} = +0.676 \times 10^{-2}$~m$^{-1}$ ($+14.38\%$, $p = 0.012$), a distribution-free placebo rank $p$-value of $\hat{p}_{\mathrm{rank}} = 0.000$, and stable leave-one-out covariate sensitivity. Together, these lines of evidence establish that the post-2019 increase in nearshore $K_d(490)$ off Morowali is a causal response to local anthropogenic forcing, not regional oceanographic variability. The null result in the control zone, where no consensus breakpoint was detected and the record remained a single regime, satisfies the critical identifying assumption of the BSTS framework: that the covariates used to construct the counterfactual were themselves unaffected by the intervention~\cite{brodersen2015}. To our knowledge, this constitutes the first satellite-derived causal attribution of coastal water clarity degradation to a specific industrial installation in the Indonesian archipelago.

The timing of the consensus breakpoint is telling when read against the IMIP expansion chronology. The breakpoint does not coincide with initial commissioning of nickel pig iron smelters in April 2015, a period during which the park operated within a comparatively modest footprint. Rather, the May 2019 breakpoint aligns with a qualitatively different phase that began in late 2018, when a multinational consortium committed to constructing high-pressure acid leaching (HPAL) facilities for battery-grade nickel in direct response to the anticipated reimposition of the complete export ban~\cite{lahadalia2024nickel,Astuti2025}. Construction of HPAL plants, together with coal-fired power generation, slag storage, and tailings infrastructure, plausibly initiated a step change in terrestrial disturbance that propagated to the marine environment within months. The steep lateritic terrain and short, high-gradient catchments along the Morowali coastline would have accelerated this transfer.

The LULC intensity analysis independently supports this interpretation. Built area expanded $3.8$-fold from $12.26$~km$^{2}$ (2017) to $46.18$~km$^{2}$ (2024), while tree cover dropped by 5.04 percentage points. The QES decomposition shows that over half of total change was exchange-type (systematic pairwise swaps between losing and gaining categories), consistent with organized, planned industrial conversion rather than stochastic fragmentation. The breakpoint may also mark the point at which cumulative footprint growth exceeded the assimilative capacity of remaining riparian and mangrove buffers, which would account for the absence of a detectable break during the 2015--2018 initial operations phase.

Several mechanistic pathways likely contribute to elevated $K_d(490)$ after May 2019. First, laterite mining involves wholesale removal of vegetation and weathered regolith from ultramafic terrain, exposing fine-grained, dispersive ferralitic soils to tropical rainfall~\cite{vanderent2013,mudd2010}. Comparable operations in New Caledonia have substantially elevated nearshore suspended sediment concentrations~\cite{david2010,dumas2010}. Second, HPAL processing, now the dominant technology at IMIP, generates large volumes of tailings containing sulfuric acid and hexavalent chromium, classified as hazardous waste under Indonesian law. Ground-level impoundment of these tailings under high tropical rainfall introduces persistent leakage risk. Third, progressive conversion of vegetated catchment to impervious surfaces increases runoff coefficients and eliminates riparian sediment trapping~\cite{walsh2005}. The relative contributions of these pathways cannot be disentangled from $K_d(490)$ alone; targeted in-situ sampling of suspended particulate matter, dissolved metals, and source-diagnostic tracers would be needed.

The ecological significance is best evaluated through the photic zone impact. The $14.38\%$ increase in $K_d(490)$ translates to euphotic zone shoaling from $97.8$ to $85.5$~m ($\Delta Z_{\mathrm{eu}} = -12.3$~m). These waters lie within the Coral Triangle, the global epicenter of marine species richness~\cite{veron2009} and a priority conservation region~\cite{asaad2018}. In oligotrophic systems, benthic communities are adapted to high baseline irradiance, and even moderate turbidity increases may therefore carry disproportionate consequences. Chronic sediment stress compromises coral photosynthesis, depletes lipid reserves, and reduces skeletal extension at concentrations below acute mortality thresholds~\cite{jones2015,fabricius2005}. Vertical reef compression (the progressive exclusion of light-dependent taxa from deeper habitats) can permanently restructure community composition~\cite{muir2015}. Laterite-derived sediment is enriched in nickel, chromium, and cobalt~\cite{dublet2012}; elevated dissolved nickel at ecologically significant concentrations has been documented in comparable systems~\cite{hedouin2007,houlbreque2011}. These combined optical and geochemical stressors threaten both marine ecosystems and coastal communities that depend on reef fisheries~\cite{hicks2019}.

The identification strategy has several strengths alongside important limitations. The BSTS framework isolates local forcing by constructing a counterfactual from unaffected control-zone observations~\cite{brodersen2015}, and the leave-one-out analysis confirms that no single covariate drives the result. Three limitations warrant emphasis. First, the 4~km GlobColour resolution integrates optical properties across areas far larger than individual plumes, attenuating the nearshore signal; our estimates are therefore conservative lower bounds. Higher-resolution sensors could resolve sub-pixel gradients but lack the deep pre-intervention record needed for stable counterfactual estimation~\cite{wang2009}. Second, residual diagnostics indicate substantial departures from Gaussianity ($g_1 = +3.86$, $g_2 = +30.00$) and significant autocorrelation, consistent with episodic turbidity pulses superimposed on a secular shift. The placebo rank test, which serves as the primary inferential basis, is distribution-free and remains valid regardless~\cite{brodersen2015}. Third, the pre-intervention impact zone exhibits a weak but significant upward trend ($\tau = 0.179$, $p < 0.001$), possibly reflecting early-phase construction, intensifying artisanal mining, or low-amplitude climatic drift. If anthropogenic, the May 2019 breakpoint marks a transition from gradual to acute degradation, a refinement that does not invalidate the causal interpretation. Finally, the absence of concurrent in-situ measurements precludes direct ground-truthing. This gap is symptomatic of Indonesia's environmental assessment framework, which does not mandate continuous marine monitoring at coastal industrial sites.

These results quantify a marine environmental externality absent from Indonesia's nickel downstreaming discourse. The progressive export bans have achieved their economic objectives~\cite{lahadalia2024nickel,Astuti2025}, but our evidence demonstrates a causally attributable marine degradation. Where comparable industries in the western Pacific accumulated coastal impacts over decades~\cite{dumas2010}, IMIP has compressed equivalent transformation into barely a decade. Counterfactual evaluations have documented substantial deforestation in Sulawesi nickel communities~\cite{lo2024}, and empirical footprint measurements have shown disturbance intensities far exceeding standard assumptions~\cite{heijlen2024}. Replication of the IMIP model at Weda Bay (North Maluku) and the Virtue Dragon complex (Southeast Sulawesi) suggests that these impacts are being reproduced wherever similar facilities are sited adjacent to oligotrophic waters.

\section{Conclusion}

This study provides satellite-derived causal evidence that rapid industrialization of Indonesia's Morowali coastline has produced a statistically detectable degradation of nearshore water clarity. $K_d(490)$ increased substantially relative to a synthetic counterfactual constructed from contemporaneous Banda Sea controls. The consensus breakpoint coincides not with initial smelter commissioning but with the onset of hyper-expansion driven by HPAL investment and the imminent export ban, a timing independently corroborated by satellite land cover data showing accelerating built-area growth and tree cover decline. The resulting euphotic zone shoaling occurs in oligotrophic waters of exceptional biodiversity value, where even moderate optical degradation can initiate cascading ecological consequences for light-dependent benthic communities. These findings expose a marine externality absent from the prevailing economic narrative surrounding Indonesia's mineral downstreaming strategy and underscore the urgent need for mandatory, continuous marine water quality monitoring at coastal industrial sites across the archipelago. The quasi-experimental framework demonstrated here, combining multi-sensor ocean color records, BSTS modeling, and distribution-free placebo testing, provides a transferable template for causal environmental impact assessment in data-limited tropical settings where conventional monitoring is absent and regulatory enforcement remains inconsistent.



\section*{Funding}
Financial support was provided by the ITB 3P Research Program (Talenta Unggul Scheme) through the Directorate of Research and Innovation, Bandung Institute of Technology (Project ID: DRI.PN-6-64-2026) and the University of California, Riverside (Dean's Distinguished Fellowship 2023).

\section*{Author Contributions}
\textbf{S. H. S. H.}: Conceptualization, Data curation, Formal analysis, Investigation, Methodology, Software, Visualization, Writing -- original draft. \textbf{A. P. H.}: Conceptualization, Investigation, Project administration, Validation, Supervision, Resources, Funding acquisition, Writing -- review \& editing. \textbf{I. P. A.}: Supervision, Resources, Writing -- review \& editing. \textbf{F. K.}: Supervision, Resources, Writing -- review \& editing. \textbf{K. A. S.}: Supervision, Resources, Writing -- review \& editing. \textbf{D. Y. W}: Conceptualization, Supervision, Writing -- review \& editing. \textbf{R. S.}: Supervision, Writing -- review \& editing. \textbf{D. E. I.}: Supervision, Funding acquisition, Writing -- review \& editing.

\section*{Data Availability}

The raw satellite and oceanographic datasets are available from their respective public repositories: ocean color and physical oceanographic variables via CMEMS (\url{https://marine.copernicus.eu}); LULC composites via Esri (\url{https://livingatlas.arcgis.com/landcoverexplorer}); bathymetry via SRTM15+V2 (\url{https://topex.ucsd.edu/WWW_html/srtm15_plus.html}). Processed datasets, scripts, and supplementary materials are available at \url{https://github.com/sandyherho/supplMorowaliOcean}.

\section*{Conflicts of Interest}

The authors declare no conflicts of interest.

\end{document}